\documentclass[12pt,a4paper]{article} %
\usepackage[a4paper, total={176mm, 240mm}, left=18mm, top=20mm]{geometry}  
\usepackage{epsfig,amsmath,amssymb,graphicx,upgreek,textcomp,revsymb}
\usepackage[english]{babel}
\usepackage{authblk}  %
\usepackage{indentfirst} %
\numberwithin{equation}{section}

\title{Thermodynamic functions of the Fermi gas \\ at arbitrary temperatures} %
\author{Yu.M. Poluektov, A.A. Soroka}  \date{}
\affil{\normalsize{National Science Center ``Kharkov Institute of Physics and Technology'', \newline 61108 Kharkov, Ukraine}} %
\affil{\small{E-mail: yuripoluektov@kipt.kharkov.ua (y.poluekt52@gmail.com)}} %

\begin{document}

\maketitle

\begin{abstract}
Thermodynamic functions of the ideal Fermi gas at arbitrary
temperatures are calculated using the standard Fermi-Stoner
functions. The properties of the Fermi-Stoner functions are
analyzed. The limiting cases of low-temperature and classical limits
with taking into account quantum corrections and the special case of
zero chemical potential are considered. \vspace{1mm}
\newline  %
{\bf Key words}: %
fermion, electron, thermodynamic functions, equation of state,
entropy, heat capacity, thermodynamic coefficients %
\vspace{1mm} \newline  %
PACS numbers:\,05.30.Fk, 05.70.Ce, 64.10.+h, 67.10.Db, 67.10.Fj, 71.10.Ca %
\end{abstract} \newpage

\tableofcontents  \vspace{10mm}

\section*{Introduction}\addcontentsline{toc}{section}{Introduction} %
The theory of ideal gases is presented in many textbooks and problem
books on statistical mechanics [1 -- 8]. However, as a rule, authors 
limit themselves to considering two limiting cases: \linebreak  %
a) high temperatures, at which the description of the motion of gas
particles on the basis of the laws of classical mechanics is valid,
and b) very low temperatures, when gases are close to degeneracy.
Meanwhile, in low-temperature investigations, a situation is quite
real when a system, which can be considered as an ideal gas (for
example, quasiparticles in quantum liquids), is no longer classical,
but is still far from strong degeneracy. In this intermediate case,
for calculations of thermodynamic functions formulas relating to the
above limiting cases cannot be used, but exact expressions should be
used. Since in the known to us literature on statistical physics the
methodology of such calculations is not presented, the current work
is devoted to the presentation of methods of calculation of
thermodynamic functions for the Fermi gas at arbitrary temperatures.
The ideal gas model allows to verify all fundamental thermodynamic
relations on the basis of a microscopic description and to express
all thermodynamic quantities through standard functions. In
addition, a detailed study of the thermodynamic properties of ideal
gases helps to better understand the general principles of
statistical mechanics.

\section{Thermodynamic functions of the Fermi gas at arbitrary temperatures}\vspace{-0mm} %
{\small\noindent Thermodynamic functions of the Fermi gas at
arbitrary temperatures are expressed in terms of the standard
Fermi-Stoner functions.}\vspace{4mm}

In the ideal Fermi gas, the momentum distribution of particles with
a half-integer spin is described by the Fermi-Dirac function
\begin{equation} \label{01}
\begin{array}{l}
\displaystyle{%
  f_k=\big(\exp\beta\xi_k + 1\big)^{-1},  %
}%
\end{array}
\end{equation}
where $\xi_k=\hbar^2k^2/2m - \mu$, $\beta=1/T$ -- the inverse
temperature, $T$ -- the temperature, $\mu$ -- the chemical
potential, $k$ -- the wave number of a particle, $m$ -- its mass.
All thermodynamic functions of the ideal Fermi gas can be expressed
through special functions defined by the formula
\begin{equation} \label{02}
\begin{array}{l}
\displaystyle{%
  \Phi_s(t)=\frac{1}{\Gamma(s)}\int_0^\infty\!\frac{z^{s-1}dz}{e^{z-t}+1}, %
}%
\end{array}
\end{equation}
where $s$ is a positive integer or half-integer number. Similar
functions were first introduced by Stoner [9] and therefore we will
call them {\it the Fermi-Stoner functions}. The parameter $t$ in
(1.2) takes any real values from $-\infty$ to $\infty$. The
properties of these functions will be discussed in detail in the next section. %

Further formulas will include the de Broglie thermal wavelength of a
particle, defined by the formula
\begin{equation} \label{03}
\begin{array}{c}
\displaystyle{\hspace{0mm}%
  \Lambda\equiv\Lambda(T)=\bigg(\frac{2\pi\hbar^2}{mT}\bigg)^{\!\!1/2}. %
}%
\end{array}
\end{equation}
As seen, the existence of the characteristic length $\Lambda$ is
conditioned by quantum laws, since the definition (1.3) contains the
Planck's constant in the numerator. When passing to the classical
description we should formally set $\hbar\rightarrow 0$, so that
$\Lambda\rightarrow 0$. In reality, the decrease of the thermal
wavelength, and hence the transition to the classical description,
occurs as the temperature increases. The criterion of transition
from the quantum to the classical description of a gas will be
discussed in more detail below.

When finding the following relations, one should use the
representation of the distribution function in the form of a series
\begin{equation} \label{04}
\begin{array}{ll}
\displaystyle{%
  \frac{1}{e^\xi + 1}=\left\{
               \begin{array}{l}
                  \displaystyle{\sum_{n=0}^\infty (-1)^n e^{n\xi},      \hspace{12.5mm} \xi <0, } \vspace{1mm}\\ %
                  \displaystyle{\sum_{n=0}^\infty (-1)^n e^{-(n+1)\xi}, \quad \xi >0. }    %
\end{array} \right.
}%
\end{array}
\end{equation}

\subsection{Thermodynamic potential}
The thermodynamic potential of the ideal Fermi gas through the
Fermi-Stoner function is expressed by the formula
\begin{equation} \label{05}
\begin{array}{l}
\displaystyle{%
  \Omega=-gT\sum_k\ln\!\big(1+e^{-\beta\xi_k}\big)=-\frac{gTV}{\Lambda^3}\,\Phi_{5/2}(t), %
}%
\end{array}
\end{equation}
where $t=\beta\mu$, $g\equiv 2s+1$, $s$ -- the particle spin. For
the electron and the nucleons $s=1/2$ and $g=2$. The parameter
$t=\beta\mu$, equal to the product of the chemical potential and the
inverse temperature, is often encountered in the theory of quantum
gases, so let us give this parameter the name -- {\it the
adiabaticity parameter}. The reason for this name will be explained
below.

\subsection{Number of particles}
The total number of particles as a function of the chemical
potential and temperature (or the parameter $t=\beta\mu$ and
temperature) is determined by the formula
\begin{equation} \label{06}
\begin{array}{l}
\displaystyle{%
  N=g\sum_k f_k=\frac{gV}{\Lambda^3}\,\Phi_{3/2}(t). %
}%
\end{array}
\end{equation}
Note that the thermodynamic relation
$N=-\big(\partial\Omega/\partial\mu\big)_T$ is fulfilled.

\subsection{Energy}
The energy is determined by the formula
\begin{equation} \label{07}
\begin{array}{l}
\displaystyle{%
  E=g\sum_k \frac{\hbar^2k^2}{2m}\,f_k=\frac{3}{2}\frac{gTV}{\Lambda^3}\,\Phi_{5/2}(t). %
}%
\end{array}
\end{equation}
As seen, the known relation $\displaystyle{\Omega=-\frac{2}{3}E}$ is fulfilled. %

\subsection{Pressure}
Since $\Omega=-pV$, the pressure is determined by the formula
\begin{equation} \label{08}
\begin{array}{l}
\displaystyle{%
  p=\frac{gT}{\Lambda^3}\,\Phi_{5/2}(t). %
}%
\end{array}
\end{equation}
Obviously, there holds the equation $\displaystyle{pV=\frac{2}{3}E}$. %

\subsection{Entropy}
The entropy of the ideal Fermi gas can be found either from the
combinatorial definition
\begin{equation} \label{09}
\begin{array}{l}
\displaystyle{%
   S=-g\sum_k\big[f_k\ln f_k + (1-f_k)\ln (1-f_k) \big], %
}
\end{array}
\end{equation}
or using the thermodynamic relation $S=-\big(\partial\Omega/\partial T\big)_\mu$: %
\begin{equation} \label{10}
\begin{array}{l}
\displaystyle{%
  S=\frac{gV}{\Lambda^3}\bigg[\frac{5}{2}\Phi_{5/2}(t)-t\Phi_{3/2}(t)\bigg]. %
}%
\end{array}
\end{equation}
Note that for the above thermodynamic quantities the known
thermodynamic relation holds $E=\mu N + TS - pV$. %

Formulas (\ref{06})\,--\,(\ref{08}), (\ref{10}) determine
parametrically (the parameter $t$) the energy, pressure and entropy
as functions of the temperature and particle number density.
Practically, however, it is sometimes more convenient to use another
dimensionless parameter $\theta$. From formulas (\ref{06}) and
(\ref{08}) it follows that
\begin{equation} \label{11}
\begin{array}{l}
\displaystyle{%
  \frac{nT}{p}\equiv\theta=\frac{\Phi_{3/2}(t)}{\Phi_{5/2}(t)}, %
}%
\end{array}
\end{equation}
where $n=N/V$ -- the particle number density. Thus, formula
(\ref{11}) uniquely links the parameters $\theta$ and $t$. The
dependence $\theta=\theta(t)$ is shown in Fig.\,1. The parameter
$\theta$ varies within the finite limits $0<\theta<1$. At
$t\rightarrow -\infty$\,\, $\theta\rightarrow 1$, and at
$t\rightarrow +\infty$\,\, $\theta\approx 5\big/2t$. By calculating
the dependence of any thermodynamic quantity on $\theta$, it is easy
to obtain its dependence on temperature, density or pressure with
the other two variables being fixed.
\begin{figure}[h!]
\vspace{-1mm}  \hspace{0mm} %
\centering
\includegraphics[width = 7.34cm]{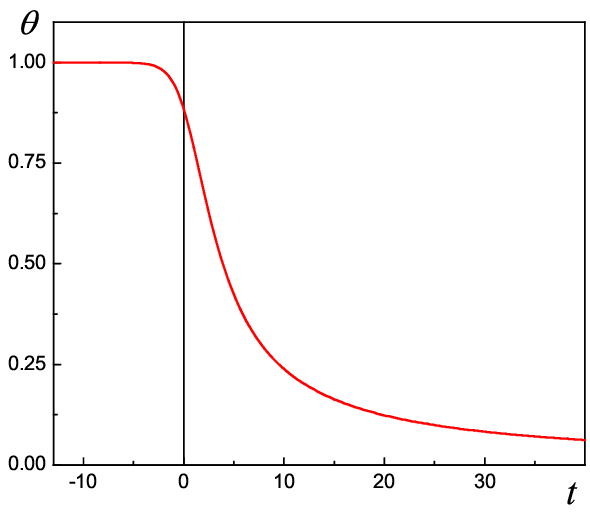} 
\vspace{-4mm} %
\caption{\label{fig01} 
Dependence of the parameter $\theta$ on the adiabaticity parameter $t$. %
}%
\end{figure}

\subsection{Adiabatic equations}
The entropy per one particle
\begin{equation} \label{12}
\begin{array}{l}
\displaystyle{%
  \sigma\equiv\frac{S}{N}=\frac{5}{2}\frac{\Phi_{5/2}(t)}{\Phi_{3/2}(t)}-t %
}%
\end{array}
\end{equation}
depends only on the parameter $t$. In adiabatic processes, when
$\sigma=const$, the parameter $t=\mu/T$ is also constant. This is
the reason for the adopted name of $t$ -- {\it the adiabaticity
parameter}. The parameter $\theta$ (1.11) also does not change in
adiabatic processes. The dependencies $\sigma=\sigma(t)$ and
$\sigma=\sigma(\theta)$ are shown in Fig.\,2. In the limit
$t\rightarrow -\infty$ the dependence (\ref{12}) becomes linear
$\sigma\cong 5/2 - t$, and at $t\rightarrow\infty$ the entropy per
one particle decreases as $\sigma\cong\pi^2/2t$. At $\theta\ll 1$
there is a linear dependence $\sigma(\theta)\approx
\frac{\pi^2}{5}\,\theta$, and at $\theta\rightarrow 1$, when
$t\approx (5/2)\ln 2 +
\ln(1-\theta)$, we have $\sigma(\theta)\approx (5/2)(1-\ln 2)-\ln(1-\theta)$. %
\begin{figure}[t!]
\vspace{-1mm}  \hspace{0mm} %
\centering
\includegraphics[width = 15cm]{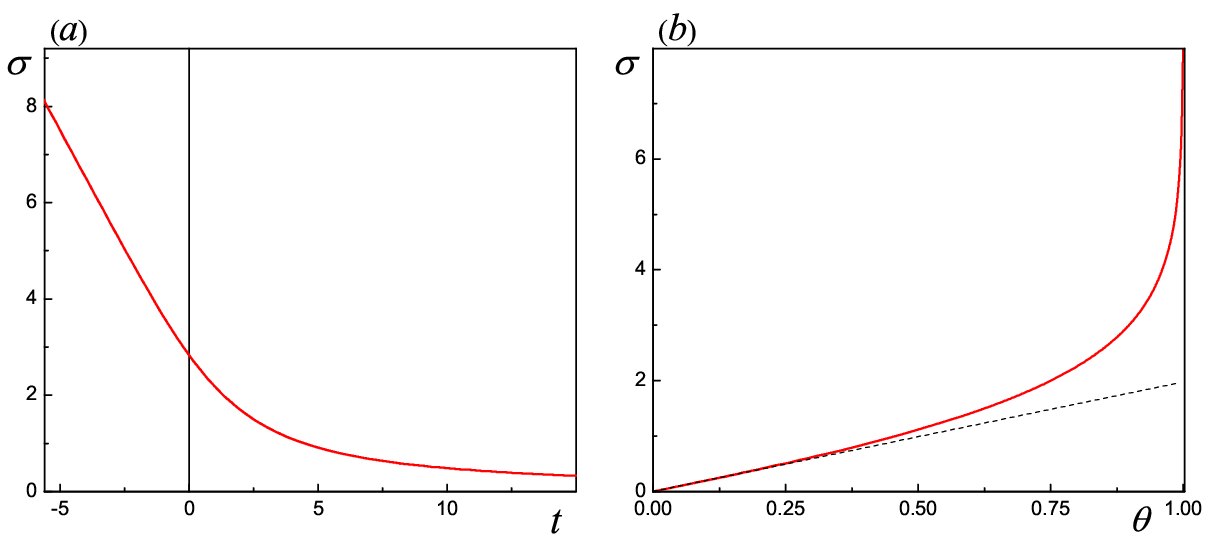} 
\vspace{-4mm} %
\caption{\label{fig02} 
Dependences of the entropy per one particle on: %
({\it a}) the parameter $t$ -- $\sigma(t)$, \newline %
({\it b}) the parameter $\theta$ -- $\sigma(\theta)$. %
}%
\end{figure}

Taking into account the constancy of $t$, from formulas (\ref{06}),
(\ref{08}) there follow the adiabatic equations
\begin{equation} \label{13}
\begin{array}{l}
\displaystyle{%
  \frac{n}{T^{3/2}}=\frac{g}{A}\,\Phi_{3/2}(t), \qquad %
  \frac{p}{n^{5/3}}=\frac{A^{2/3}}{g^{2/3}}\frac{\Phi_{5/2}(t)}{\Phi_{3/2}^{5/3}(t)}, \qquad %
  \frac{p}{T^{5/2}}=\frac{g}{A}\,\Phi_{5/2}(t),  %
}%
\end{array}
\end{equation}
where $A=\big(2\pi\hbar^2/m\big)^{3/2}$. The values of the constant
quantities on the right-hand side of equations (\ref{13}) are
completely determined by the adiabaticity parameter $t$ or the
parameter $\theta$.

\subsection{The degree of ``quantumness'' of the gas (``quantumness'')}
Depending on the particle number density and temperature, the
quantum mechanical properties of the gas will manifest themselves to
a greater or lesser extent. As a measure of ``quantumness'' it is
natural to use the ratio of the de Broglie thermal wavelength
(\ref{03}) to the average distance between particles $l=n^{-1/3}$. %
However, it is more convenient to take as such a measure the ratio
of the volumes of spheres with radii $\Lambda$ and $l$. Thus, the
number $q$, characterizing the measure of ``quantumness'' of the
gas, is defined as the ratio
\begin{equation} \label{14}
\begin{array}{l}
\displaystyle{%
  q\equiv\bigg(\frac{\Lambda}{l}\bigg)^{\!3}=\Lambda^3n=g\Phi_{3/2}(t). %
}%
\end{array}
\end{equation}
As seen, $q$ depends only on the adiabaticity parameter $t$, i.e.
the value of this parameter (or $\theta$) uniquely determines the
degree of ``quantumness'' of the gas. In the following, for brevity
we will call the number $q$ ``quantumness''. With decreasing
temperature at a fixed density the thermal length increases, and
hence the quantumness also increases in this case. The dependences
of the quantumness $q$ on the parameters $t$ and $\theta$ are
presented in Fig.\,3. The quantumness of the gas increases with
increasing the parameter $t$ and with decreasing $\theta$. In the
limit $t\rightarrow -\infty$ or $\theta\rightarrow 1$, the
quantumness $q\rightarrow 0$ and the motion of gas particles obeys
the laws of classical mechanics. %
The quantumness can be found based on the temperature and the
particle number density, and the adiabaticity parameter $t$\, for a
given state of the gas is determined based on the known quantumness
from formula (\ref{14}).
\begin{figure}[h!]
\vspace{00mm}  \hspace{0mm} %
\centering
\includegraphics[width = 14.9cm]{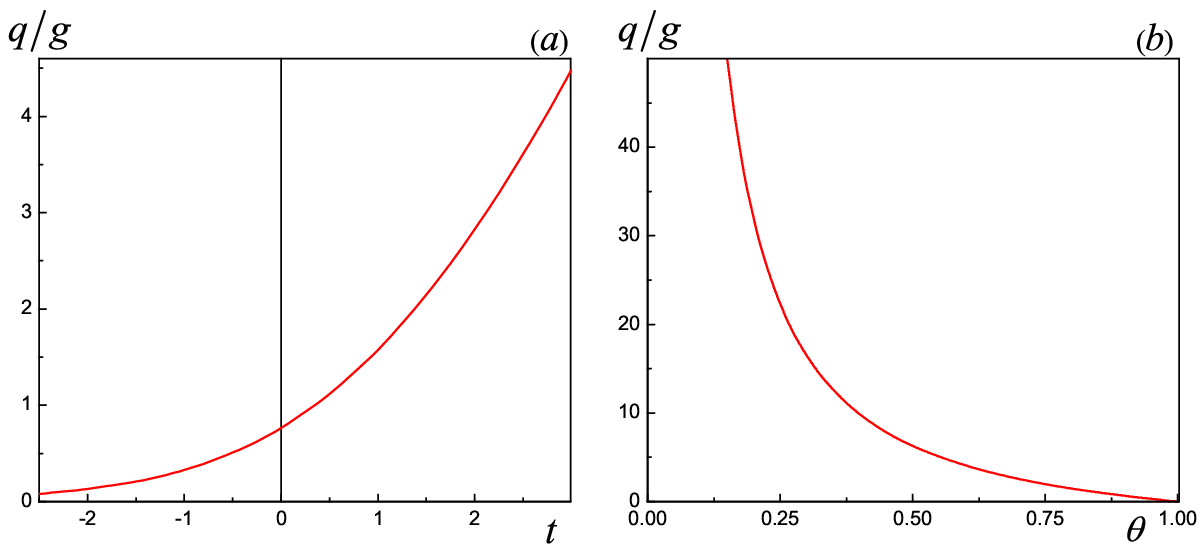} 
\vspace{-4mm} %
\caption{\label{fig03} 
Dependences of the ``quantumness'' on: %
({\it a}) the parameter $t$ -- $q(t)/g$, \newline %
({\it b}) the parameter $\theta$ -- $q(\theta)/g$. %
}%
\end{figure}

\subsection{Heat capacities}
When finding heat capacities and thermodynamic coefficients, it is
convenient to use the following expressions for differentials
\begin{equation} \label{15}
\begin{array}{l}
\displaystyle{%
  dN=\frac{N}{V}\,dV + \frac{3N}{2T}\,dT + \frac{gV}{\Lambda^3}\,\Phi_{1/2}\,dt, %
}%
\end{array}
\end{equation}
\vspace{00mm}
\begin{equation} \label{16}
\begin{array}{l}
\displaystyle{%
  dp=\frac{5p}{2T}\,dT + \frac{gT}{\Lambda^3}\,\Phi_{3/2}\,dt, %
}%
\end{array}
\end{equation}
\vspace{00mm}
\begin{equation} \label{17}
\begin{array}{l}
\displaystyle{%
  dS=\frac{S}{V}\,dV + \frac{3S}{2T}\,dT + \frac{gV}{\Lambda^3}\bigg(\frac{3}{2}\,\Phi_{3/2}-t\Phi_{1/2}\bigg)dt. %
}%
\end{array}
\end{equation}
Here, the relation is used $d\Lambda/\Lambda=-dT/2T$. The heat
capacity for an arbitrary change of thermodynamic variables with
temperature is defined by the expression
\begin{equation} \label{18}
\begin{array}{l}
\displaystyle{%
  C\equiv T\frac{dS}{dT}=\frac{ST}{V}\,\frac{dV}{dT} + \frac{3S}{2} + \frac{gVT}{\Lambda^3}\bigg(\frac{3}{2}\,\Phi_{3/2}-t\Phi_{1/2}\bigg)\frac{dt}{dT}. %
}%
\end{array}
\end{equation}
As a rule, systems with a fixed average number of particles are
considered. In this case $dN=0$, and from (\ref{15}) it follows that
\begin{equation} \label{19}
\begin{array}{l}
\displaystyle{%
  \frac{dV}{dT}=-\frac{3V}{2T} - \frac{gV^2}{N\Lambda^3}\,\Phi_{1/2}\frac{dt}{dT}. %
}%
\end{array}
\end{equation}
For a constant volume from (\ref{19}) we have
\begin{equation} \label{20}
\begin{array}{l}
\displaystyle{%
  \frac{dt}{dT}=-\frac{3}{2}\frac{N\Lambda^3}{gVT}\,\Phi_{1/2}^{\,-1}, %
}%
\end{array}
\end{equation}
and for a constant pressure from (\ref{16}) it follows
\begin{equation} \label{21}
\begin{array}{l}
\displaystyle{%
  \frac{dt}{dT}=-\frac{5}{2}\frac{p\Lambda^3}{gT^2}\,\Phi_{3/2}^{\,-1}. %
}%
\end{array}
\end{equation}
Formulas (\ref{18})\,--\,(\ref{21}) allow us to obtain formulas for
the heat capacities of the gas under certain conditions. The heat
capacity per one particle at a constant volume is determined by the formula %
\begin{equation} \label{22}
\begin{array}{l}
\displaystyle{%
  c_V\equiv\frac{C_V}{N}=\frac{15}{4}\frac{\Phi_{3/2}(t)}{\Phi_{1/2}(t)}\bigg[\frac{\Phi_{1/2}(t)\Phi_{5/2}(t)}{\Phi_{3/2}^2(t)}-\frac{3}{5}\bigg], %
}%
\end{array}
\end{equation}
and the heat capacity at a constant pressure -- by the formula
\begin{equation} \label{23}
\begin{array}{l}
\displaystyle{%
  c_p\equiv\frac{C_p}{N}=\frac{25}{4}\frac{\Phi_{5/2}(t)}{\Phi_{3/2}(t)}\bigg[\frac{\Phi_{1/2}(t)\Phi_{5/2}(t)}{\Phi_{3/2}^2(t)}-\frac{3}{5}\bigg]. %
}%
\end{array}
\end{equation}
Formulas (\ref{22}),\,(\ref{23}), together with (\ref{11}),
determine parametrically the dependences of the heat capacities on
the positive parameter $\theta$: $C_V/N=c_V(\theta)$, $C_p/N=c_p(\theta)$. %
The dependences of the heat capacities and their ratios on the
parameters $t$ and $\theta$ are shown in Fig.\,4. In the classical
limit $t\rightarrow -\infty$, the heat capacities tend to constant
values $c_p=5/2$, $c_V=3/2$, and their ratio to $c_p/c_V=5/3$. In
the quantum limit $t\rightarrow\infty$, the heat capacities tend to
zero according to the law $c_p\cong c_V\cong\pi^2/2t$, and their
ratio tends to unity. For small $\theta\ll 1$ (the quantum limit) we
have $c_V(\theta)=c_p(\theta)\approx\frac{\pi^2}{5}\,\theta$.
\begin{figure}[h!]
\vspace{-02mm}  \hspace{0mm} %
\centering
\includegraphics[width = 14.9cm]{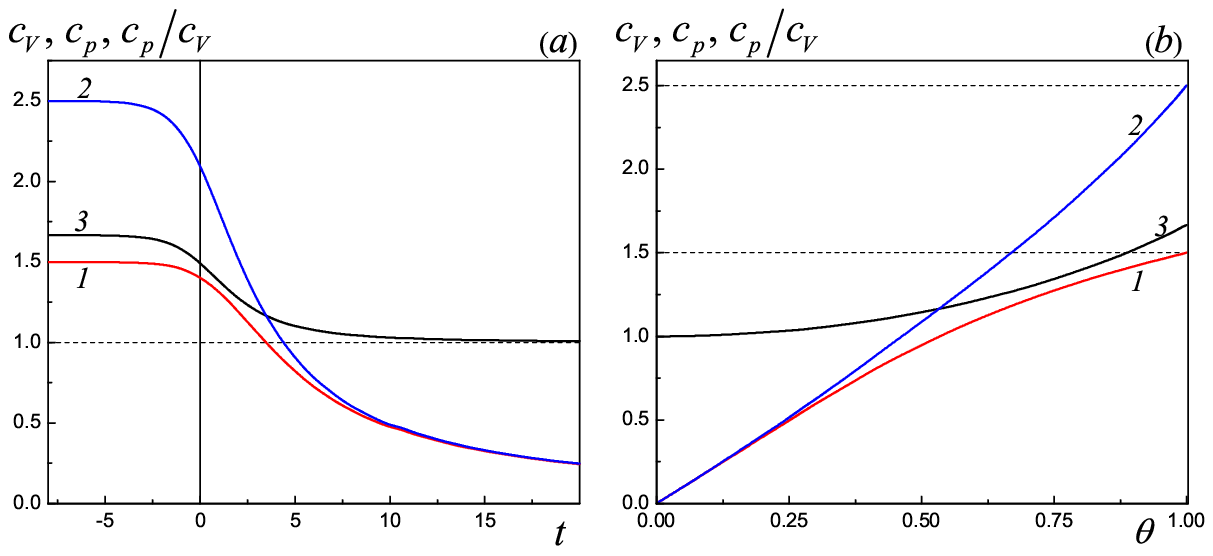} 
\vspace{-4mm} %
\caption{\label{fig04} 
Dependences of the heat capacities per one particle and their ratios
on the parameters $t$ and $\theta$: %
({\it a}) {\it 1} -- $c_V(t)$, {\it 2} -- $c_p(t)$, {\it 3} -- $c_p(t)/c_V(t)$;  %
({\it b}) {\it 1} -- $c_V(\theta)$, {\it 2} -- $c_p(\theta)$, {\it 3} -- $c_p(\theta)/c_V(\theta)$. %
}%
\end{figure}

\noindent Let us also give formulas for the ratio and the difference
of heat capacities
\begin{equation} \label{24}
\begin{array}{l}
\displaystyle{%
  \frac{C_p}{C_V}=\frac{5}{3}\frac{\Phi_{1/2}(t)\Phi_{5/2}(t)}{\Phi_{3/2}^2(t)}, %
}%
\end{array}
\end{equation}
\begin{equation} \label{25}
\begin{array}{l}
\displaystyle{%
  C_p-C_V=\frac{25}{4}\frac{gV}{\Lambda^3}\frac{\Phi_{3/2}^2(t)}{\Phi_{1/2}(t)}\bigg[\frac{\Phi_{1/2}(t)\Phi_{5/2}(t)}{\Phi_{3/2}^2(t)}-\frac{3}{5}\bigg]^{\!2}. %
}%
\end{array}
\end{equation}
From (\ref{25}) it is obvious that the condition $C_p-C_V > 0$ is
satisfied. The heat capacities per one particle at constant values
of volume and pressure (\ref{22}),\,(\ref{23}), as well as the ratio
of heat capacities (\ref{24}), are functions of only the
adiabaticity parameter and remain constant in adiabatic processes.

\subsection{Isothermal and adiabatic speeds of sound}
The squares of isothermal and adiabatic speeds of sound are defined
as derivatives of the pressure with respect to the density under the
appropriate conditions and have the form
\begin{equation} \label{26}
\begin{array}{l}
\displaystyle{%
  u_T^2=\frac{1}{m}\bigg(\frac{\partial p}{\partial n}\bigg)_{\!T}=\frac{T}{m}\frac{\Phi_{3/2}(t)}{\Phi_{1/2}(t)}, %
}%
\end{array}
\end{equation}
\begin{equation} \label{27}
\begin{array}{l}
\displaystyle{%
  u_\sigma^2=\frac{1}{m}\bigg(\frac{\partial p}{\partial n}\bigg)_{\!\sigma}=\frac{5T}{3m}\frac{\Phi_{5/2}(t)}{\Phi_{3/2}(t)}. %
}%
\end{array}
\end{equation}
Their ratio is equal to the ratio of heat capacities (\ref{24})
\begin{equation} \label{28}
\begin{array}{l}
\displaystyle{%
  \frac{u_\sigma^2}{u_T^2}=\frac{C_p}{C_V}=\frac{5}{3}\frac{\Phi_{1/2}(t)\Phi_{5/2}(t)}{\Phi_{3/2}^2(t)}. %
}%
\end{array}
\end{equation}
We emphasize that this relation is valid at all temperatures.

It is of interest to determine the average of the square of the
momentum of a gas particle. Based on the definition of the average,
and taking into account relations (\ref{06}),\,(\ref{07}), we have:
\begin{equation} \label{29}
\begin{array}{l}
\displaystyle{%
  \big\langle\hbar^2k^2\big\rangle=\frac{g\sum_k \hbar^2k^2f_k}{g\sum_k f_k}=\frac{2mE}{N}=3mT\,\frac{\Phi_{5/2}(t)}{\Phi_{3/2}(t)}. %
}%
\end{array}
\end{equation}
From here it follows that the average of the square of the velocity
of a gas particle is proportional to the square of the adiabatic
speed of sound (\ref{27}) for all values of thermodynamic variables:
\begin{equation} \label{30}
\begin{array}{l}
\displaystyle{%
  \big\langle\upsilon^2\big\rangle=\frac{\big\langle\hbar^2k^2\big\rangle}{m^2}=\frac{9}{5}\,u_\sigma^2. %
}%
\end{array}
\end{equation}

Since the functions $\Phi_s(t)$ take only positive values, the
squares of the sound speeds are positive and %
{\it the conditions of thermodynamic stability}
\begin{equation} \label{31}
\begin{array}{l}
\displaystyle{%
  \bigg(\frac{\partial p}{\partial n}\bigg)_{\!T} > 0, \qquad C_V>0, %
}%
\end{array}
\end{equation}
or
\begin{equation} \label{32}
\begin{array}{l}
\displaystyle{%
  \bigg(\frac{\partial p}{\partial n}\bigg)_{\!\sigma} > 0, \qquad C_p>0, %
}%
\end{array}
\end{equation}
are satisfied for the ideal Fermi gas. In addition, the pressure is
always positive $p>0$.

\vspace{-1mm}
\subsection{Number of particles collisions with a wall} \vspace{-1mm}%
Above, the formula for pressure (\ref{08}) was found using the
thermodynamic relation. It is useful to determine the gas pressure
on a wall directly through the number of particle collisions. Let
the gas fill the space at $z>0$. The number of impacts per unit area
of a wall per unit time from particles with momenta in the range
from $\hbar{\bf k}$ to $\hbar{\bf k}+\hbar d{\bf k}$ is equal to
\begin{equation} \label{33}
\begin{array}{l}
\displaystyle{%
  d\nu=g|\upsilon_z|f_k\frac{2\pi\kappa\,d\kappa\, dk_z}{(2\pi)^3}, %
}%
\end{array}
\end{equation}
where $|\upsilon_z|=\hbar|k_z|/m$, $\kappa=\sqrt{k_x^2+k_y^2}$. The
total number of impacts per unit area of a wall per unit time is
obtained by integrating the relation (\ref{33}):
\begin{equation} \label{34}
\begin{array}{l}
\displaystyle{%
  \nu=\frac{gmT^2}{4\pi^2\hbar^3}\,\Phi_2(t). %
}%
\end{array}
\end{equation}
Since at each collision a particle transfers momentum $2m\upsilon_z$
to a wall, then, multiplying (\ref{33}) by this quantity and
integrating, we obtain formula (\ref{08}) for the pressure. Note
that the quantities characterizing the volumetric properties of the
gas are expressed through functions $\Phi_s(t)$ with a half-integer
index $s$, while the number of impacts per unit area of a wall
(\ref{34}) is expressed through the function with an integer $s$.

\begin{figure}[b!]
\vspace{-00mm}  \hspace{0mm} %
\centering
\includegraphics[width = 15.0cm]{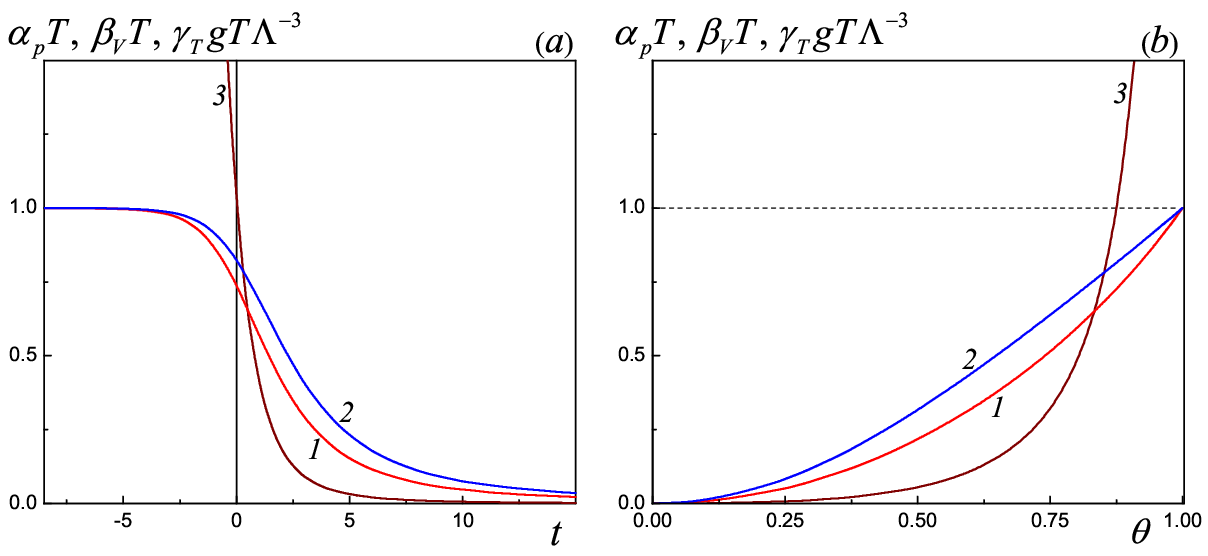} 
\vspace{-4mm} %
\caption{\label{fig05} 
Dependencies of thermodynamic coefficients on: \newline %
({\it a}) $t$:  {\it 1} -- $\alpha_p(t)T$, {\it 2} -- $\beta_V(t)T$, {\it 3} -- $\gamma_T(t)gT\Lambda^{-3}$;  %
({\it b}) $\theta$:  {\it 1} -- $\alpha_p(\theta)T$, {\it 2} -- $\beta_V(\theta)T$, {\it 3} -- $\gamma_T(\theta)gT\Lambda^{-3}$.  %
}%
\end{figure}

\subsection{Thermodynamic coefficients}
Let us also present following from (\ref{15}),\,(\ref{16}) formulas
for {\it the coefficient of volumetric expansion}
\begin{equation} \label{35}
\begin{array}{l}
\displaystyle{%
  \alpha_p=\frac{1}{V}\bigg(\frac{\partial V}{\partial T}\bigg)_{\!p}=\frac{5}{2T}\bigg[\frac{\Phi_{1/2}(t)\Phi_{5/2}(t)}{\Phi_{3/2}^2(t)}-\frac{3}{5}\bigg], %
}%
\end{array}
\end{equation}
{\it the isothermal compressibility}
\begin{equation} \label{36}
\begin{array}{l}
\displaystyle{%
  \gamma_T=-\frac{1}{V}\bigg(\frac{\partial V}{\partial p}\bigg)_{\!T}=\frac{\Lambda^3}{gT}\frac{\Phi_{1/2}(t)}{\Phi_{3/2}^2(t)} %
}%
\end{array}
\end{equation}
and {\it the isochoric thermal pressure coefficient}
\begin{equation} \label{37}
\begin{array}{l}
\displaystyle{%
  \beta_V=\frac{1}{p}\bigg(\frac{\partial p}{\partial T}\bigg)_{\!V}= %
  \frac{5}{2T}\frac{\Phi_{3/2}^2(t)}{\Phi_{1/2}(t)\Phi_{5/2}(t)}\bigg[\frac{\Phi_{1/2}(t)\Phi_{5/2}(t)}{\Phi_{3/2}^2(t)}-\frac{3}{5}\bigg]. %
}%
\end{array}
\end{equation}
Comparing with (\ref{25}), it is easy to verify that the known
thermodynamic relation is fulfilled
\begin{equation} \label{38}
\begin{array}{l}
\displaystyle{%
  C_p-C_V = TV\frac{\alpha_p^2}{\gamma_T}. %
}%
\end{array}
\end{equation}
The remaining thermodynamic coefficients that can be constructed
from the variables $T,S$ and $p,V$ are expressed through the above
coefficients and the heat capacities \cite{RR}. The dependences of
thermodynamic coefficients (\ref{35})\,--\,(\ref{37}) on the
parameters $t$ and $\theta$ are shown in Fig.\,5.

\section{Properties of the Fermi-Stoner functions}\vspace{-0mm} %
{\small\noindent The general properties of the Fermi-Stoner
functions and their calculation in the high-temperature and
low-temperature limits are considered.}\vspace{4mm}

In this section we consider in more detail the properties of the
{\it Fermi-Stoner} functions defined by the formula
\begin{equation} \label{2.1}
\begin{array}{l}
\displaystyle{%
  \Phi_s(t)=\frac{1}{\Gamma(s)}\int_0^\infty\!\frac{z^{s-1}dz}{e^{z-t}+1}, %
}%
\end{array}
\end{equation}
where $s$ can be a positive integer or half-integer number,
$\Gamma(s)$ is the gamma function, $t\equiv\mu/T$. As a rule, it is
sufficient to use functions with indices $s=1/2, 3/2, 5/2$, as well
as the derivative $\Phi_{1/2}'(t)$. In such case:
$\Gamma(1/2)=\sqrt{\pi}$, $\Gamma(3/2)=\sqrt{\pi}/2$,
$\Gamma(5/2)=3\sqrt{\pi}/4$. The form of these functions is shown in Fig.\,6. %
\begin{figure}[h!]
\vspace{-02mm}  \hspace{0mm} %
\centering
\includegraphics[width = 7.16cm]{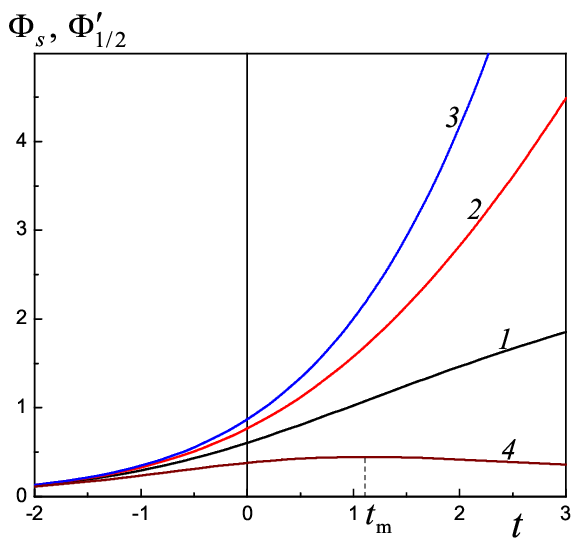} 
\vspace{-4mm} %
\caption{\label{fig06} 
Graphs of the functions $\Phi_s(t)$ at $s=1/2, 3/2, 5/2$ and the derivative $\Phi_{1/2}'(t)$; \newline %
the maximum of the function $\Phi_{1/2}'(t)$ lies at $t_m=1.11$. %
}%
\end{figure}

At $t\le 0$ the Stoner functions can be represented as a series
\begin{equation} \label{2.2}
\begin{array}{l}
\displaystyle{%
  \Phi_s(t)=\sum_{n=1}^{\infty}(-1)^{n+1}\frac{e^{nt}}{n^s}. %
}%
\end{array}
\end{equation}
At positive values of the parameter $t$ the functions (\ref{2.1})
can be written in the form
\begin{equation} \label{2.3}
\begin{array}{l}
\displaystyle{%
  \Phi_s(t)=\frac{2t^s}{\Gamma(s)}\int_0^\infty\frac{x^{2s-1}dx}{e^{\,t(x^2-1)}+1}. %
}%
\end{array}
\end{equation}
In particular,
\begin{equation} \label{2.4}
\begin{array}{l}
\displaystyle{%
  \Phi_0(t)=\frac{e^t}{1+e^t}, \qquad \Phi_1(t)=\ln\big(1+e^t\big). %
}%
\end{array}
\end{equation}

For $s\ge 1$, the derivative of the function of order $s$ is
determined by the function of  order $s-1$:
\begin{equation} \label{2.5}
\begin{array}{l}
\displaystyle{%
  \frac{d\Phi_s(t)}{dt}=\Phi_{s-1}(t). %
}%
\end{array}
\end{equation}
The functions $\Phi_s(t)$ are positive, and
\begin{equation} \label{2.6}
\begin{array}{l}
\displaystyle{%
  \Phi_s(t) > \Phi_{s-1}(t). %
}%
\end{array}
\end{equation}
At the characteristic point $t=0$ the formula is valid
\begin{equation} \label{2.7}
\begin{array}{l}
\displaystyle{%
  \Phi_s(0)=\sum_{n=1}^{\infty}\frac{(-1)^{n+1}}{n^s}=\big(1-2^{1-s}\big)\zeta(s), %
}%
\end{array}
\end{equation}
where $\zeta(s)$ is the Riemann zeta function \cite{AS}. Taking into
account the values
$$ \zeta(1/2)=-1.46035, \qquad \zeta(3/2)=2.61238, \qquad \zeta(5/2)=1.34149, $$
we obtain
$$ \Phi_{1/2}(0)=0.60490, \qquad \Phi_{3/2}(0)=0.76515, \qquad \Phi_{5/2}(0)=0.86720. $$

Since formula (\ref{2.5}) is not valid for the derivative
$\Phi_{1/2}'(t)$, this and higher derivatives of the function
$\Phi_{1/2}(t)$ should be calculated separately. At $t<0$, the
derivative $\Phi_{1/2}'(t)$ can be calculated by differentiating the
series (\ref{2.2}), so that
\begin{equation} \label{2.8}
\begin{array}{l}
\displaystyle{%
  \Phi_{1/2}'(t)=\sum_{n=1}^{\infty}\,(-1)^{n+1}\sqrt{n}\,e^{nt}.  %
}%
\end{array}
\end{equation}
At $t<0$ the higher derivatives are calculated similarly. The series
(\ref{2.8}) converges the faster the smaller the value of $t$.

To calculate the derivatives of the function $\Phi_{1/2}(t)$ for
arbitrary $t$, including $t\ge 0$, let us define the integrals
\begin{equation} \label{2.9}
\begin{array}{l}
\displaystyle{%
  \Phi_{1/2}(t;n)\equiv\frac{1}{\sqrt{\pi}}\int_0^{\infty}\frac{z^{-1/2}\,dz}{\big(e^{z-t}+1\big)^n} =  %
  \frac{2}{\sqrt{\pi}}\int_0^{\infty}\frac{du}{\big(e^{\,u^{\!2}-t}+1\big)^n}\,.  %
}%
\end{array}
\end{equation}
The second form of recording the integral is more convenient for
numerical calculations due to its faster convergence. Obviously,
$\Phi_{1/2}(t; 1)=\Phi_{1/2}(t)$. The derivatives of the function
$\Phi_{1/2}(t; n)$ with respect to $t$ are expressed through the
integrals (\ref{2.9}):
\begin{equation} \label{2.10}
\begin{array}{l}
\displaystyle{%
  \frac{d\Phi_{1/2}(t;n)}{dt}=n\big[\Phi_{1/2}(t;n)-\Phi_{1/2}(t;n+1)\big], %
}%
\end{array}
\end{equation}
\vspace{00mm}
\begin{equation} \label{2.11}
\begin{array}{l}
\displaystyle{%
  \frac{d^2\Phi_{1/2}(t;n)}{dt^2}=n\big[n\Phi_{1/2}(t;n)-(2n+1)\Phi_{1/2}(t;n+1)+(n+1)\Phi_{1/2}(t;n+2)\big], %
}%
\end{array}
\end{equation}
\vspace{00mm}
\begin{equation} \label{2.12}
\begin{array}{l}
\displaystyle{%
  \frac{d^3\Phi_{1/2}(t;n)}{dt^3}=n\big[n^2\Phi_{1/2}(t;n)-(3n^2+3n+1)\Phi_{1/2}(t;n+1)+ %
}\vspace{3mm}\\ %
\displaystyle{\hspace{22mm}%
  +3(n+1)^2\Phi_{1/2}(t;n+2)-(n+1)(n+2)\Phi_{1/2}(t;n+3)\big],   %
}%
\end{array}
\end{equation}
\vspace{00mm}
\begin{equation} \label{2.13}
\begin{array}{l}
\displaystyle{%
  \frac{d^4\Phi_{1/2}(t;n)}{dt^4}=n\big[n^3\Phi_{1/2}(t;n)-(4n^3+6n^2+4n+1)\Phi_{1/2}(t;n+1)+ %
}\vspace{3mm}\\ %
\displaystyle{\hspace{12mm}%
  +(n+1)(6n^2+12n+7)\Phi_{1/2}(t;n+2)-(n+1)(n+2)(4n+6)\Phi_{1/2}(t;n+3) + %
}\vspace{3mm}\\ %
\displaystyle{\hspace{23mm}%
  +(n+1)(n+2)(n+3)\Phi_{1/2}(t;n+4)\big].  %
}%
\end{array}
\end{equation}
At $n=1$, the relations (\ref{2.10})\,--\,(\ref{2.13}) determine the
derivatives of the function $\Phi_{1/2}(t)$. In particular, at $t=0$ we have %
$$ \Phi_{1/2}'(0)=0.38011, \quad  \Phi_{1/2}''(0)=0.11868, \quad \Phi_{1/2}'''(0)=-0.087841, \quad \Phi_{1/2}^{({\rm IV})}(0)=-0.096048. $$ %

In the limit $t\rightarrow -\infty$, which corresponds to the
classical case, it is sufficient to restrict ourselves to the first
terms of the series (\ref{2.2})
\begin{equation} \label{2.14}
\begin{array}{l}
\displaystyle{%
  \Phi_s(t)\approx e^t - \frac{e^{2t}}{2^s} + \frac{e^{3t}}{3^s}  - \ldots\,. %
}%
\end{array}
\end{equation}
In this approximation the relation of $t$ with the parameter
$\theta=nT/p$ (\ref{11}) is given by the formulas
\begin{equation} \label{2.15}
\begin{array}{l}
\displaystyle{%
 \theta-1= -\frac{e^t}{2^{5/2}} + \bigg(\frac{2}{3^{5/2}}-\frac{1}{32}\bigg)e^{2t}, %
}%
\end{array}
\end{equation}
\vspace{00mm}
\begin{equation} \label{2.16}
\begin{array}{l}
\displaystyle{%
  e^t=2^{5/2}(1-\theta)+2^{17/2}\bigg(\frac{1}{3^{5/2}}-\frac{1}{64}\bigg)(1-\theta)^2. %
}%
\end{array}
\end{equation}

Another limiting case $t\rightarrow +\infty$ corresponds to the case
of a quantum gas close to degeneracy near zero temperature. In this
case, there holds an asymptotic expansion that is valid up to
exponentially small terms
\begin{equation} \label{2.17}
\begin{array}{l}
\displaystyle{%
  \Phi_s(t)=\frac{t^s}{\Gamma(s)}\bigg[\frac{1}{s}+2\sum_{k=0}^\infty %
  C_{s-1}^{2k+1}\frac{(2k+1)!}{t^{2k+2}}\big(1-2^{-(2k+1)}\big)\zeta(2k+2)\bigg]. %
}%
\end{array}
\end{equation}
Here for the binomial coefficients at integer $\nu\equiv n$ we have %
$C_\nu^r\equiv C_n^r=\frac{n!}{(n-r)!\,r!}$. 
If $\nu$ is not an integer, then $C_\nu^r=\frac{\nu(\nu-1)\,\ldots\,
(\nu-r+1)}{r!}$. The first terms of the expansion (\ref{2.17}) have
the form
\begin{equation} \label{2.18}
\begin{array}{l}
\displaystyle{%
  \Phi_s(t)=\frac{t^s}{s\Gamma(s)}\bigg[1+\frac{\pi^2}{6}\frac{s(s-1)}{t^2}+\frac{7\pi^4}{360}\frac{s(s-1)(s-2)(s-3)}{t^4}+\ldots\bigg]. %
}%
\end{array}
\end{equation}
In particular,
\begin{equation} \label{2.19}
\begin{array}{l}
\displaystyle{%
  \Phi_{1/2}(t)=\frac{2t^{1/2}}{\sqrt{\pi}}\bigg[1-\frac{\pi^2}{24\,t^2}-\frac{7\pi^4}{384\,t^4}\bigg], %
}%
\end{array}
\end{equation}
\vspace{00mm}
\begin{equation} \label{2.20}
\begin{array}{l}
\displaystyle{%
  \Phi_{3/2}(t)=\frac{4t^{3/2}}{3\sqrt{\pi}}\bigg[1+\frac{\pi^2}{8\,t^2}+\frac{7\pi^4}{640\,t^4}\bigg], %
}%
\end{array}
\end{equation}
\vspace{00mm}
\begin{equation} \label{2.21}
\begin{array}{l}
\displaystyle{%
  \Phi_{5/2}(t)=\frac{8t^{5/2}}{15\sqrt{\pi}}\bigg[1+\frac{5\pi^2}{8\,t^2}-\frac{7\pi^4}{384\,t^4}\bigg], %
}%
\end{array}
\end{equation}
\vspace{00mm}
\begin{equation} \label{2.22}
\begin{array}{l}
\displaystyle{%
  \Phi_{2}(t)=\frac{t^2}{2}\bigg[1+\frac{\pi^2}{3\,t^2}\bigg]. %
}%
\end{array}
\end{equation}
In this approximation the parameters $t$ and $\theta$ are linked by
the relations
\begin{equation} \label{2.23}
\begin{array}{l}
\displaystyle{%
  \theta=\frac{5}{2t}\bigg[1-\frac{\pi^2}{2t^2}+\frac{41\pi^4}{120\,t^4}\bigg], %
}%
\end{array}
\end{equation}
\vspace{-01mm}
\begin{equation} \label{2.24}
\begin{array}{l}
\displaystyle{%
  \frac{1}{t}=\frac{2}{5}\,\theta + \frac{4\pi^2}{125}\,\theta^3. %
}\vspace{01mm}%
\end{array}
\end{equation}
The derivation of the asymptotic expansion (\ref{2.17}) with account
of exponential terms is given in Appendix A.

\section{Thermodynamic quantities in the limiting and special cases}\vspace{-0mm} %

{\small\noindent There is considered the behavior of thermodynamic
quantities in the classical limit with account of the quantum
correction, in the low-temperature limit and at zero chemical
potential.}\vspace{0mm}

\subsection{The classical limit and quantum corrections}
Let us consider the limiting case $t\rightarrow -\infty$. In this
case, it is sufficient to take into account some first terms in the
series (\ref{2.2}). We restrict ourselves to taking into account the
first two terms, the first of which gives the classical limit, and
the second one gives the main quantum correction to it. With a
chosen accuracy the parameter $t$ can be expressed from formulas
(\ref{14}),\,(\ref{2.14}) through the parameter $q=\Lambda^3n$,
which was called the quantumness:
\begin{equation} \label{3.01}
\begin{array}{l}
\displaystyle{%
  e^t=\frac{q}{g}\bigg(1+\frac{q}{2^{3/2}g}\bigg). %
}%
\end{array}
\end{equation}
This expansion is valid under condition $q\ll 1$, which is satisfied
if the de Broglie thermal wavelength of a particle is much smaller
than the average distance between gas particles.

The energy and pressure of the gas in this limit are given by the formulas %
\begin{equation} \label{3.02}
\begin{array}{l}
\displaystyle{%
  E=\frac{3NT}{2}\bigg(1+\frac{q}{2^{5/2}g}\bigg), %
}%
\end{array}
\end{equation}
\vspace{00mm}
\begin{equation} \label{3.03}
\begin{array}{l}
\displaystyle{%
  p=nT\bigg(1+\frac{q}{2^{5/2}g}\bigg). %
}%
\end{array}
\end{equation}
As seen, the quantum correction to the ideal Fermi gas leads to an
increase of the pressure, which corresponds to the effective
repulsion of particles. This effect is conditioned by the
antisymmetry of the wave function of a system of identical Fermi
particles with respect to permutations.

The parameter $t$ enters into the entropy linearly, so that
\begin{equation} \label{3.04}
\begin{array}{l}
\displaystyle{%
  t=\ln\!\bigg(\frac{q}{g}\bigg) + \frac{q}{2^{3/2}g}\,. %
}%
\end{array}
\end{equation}
The entropy takes the form $S=S_{cl}+S'$, where entropy in the
classical limit is determined by the formula
\begin{equation} \label{3.05}
\begin{array}{c}
\displaystyle{%
  S_{cl}=N\ln\!\bigg(\frac{ge^{5/2}}{q}\bigg).%
}%
\end{array}
\end{equation}
This formula is called the Sakura-Tetrode formula \cite{HT,OS}. As
seen, the entropy per one particle $S_{cl}/N$ is determined only by
the quantumness $q$. Note that this formula in the high-temperature
limit includes the Planck's constant $\hbar$ (through the thermal
length). This indicates the special nature of this thermodynamic
function, which even in the classical limit requires attracting the
quantum-mechanical concepts for its correct definition. The
mentioned circumstance is caused by the combinatorial nature of
entropy, because of which its calculation requires taking into
account the quantum-mechanical indistinguishability of particles at
all temperatures. This also indicates that statistical physics can
be built consistently only on the basis of quantum mechanics. The
next quantum mechanical correction to the entropy has the form
\begin{equation} \label{3.06}
\begin{array}{c}
\displaystyle{%
  S'=\frac{Nq}{2^{7/2}g}. %
}%
\end{array}
\end{equation}

The heat capacities at constant volume and pressure with account of
quantum corrections are expressed by the formulas
\begin{equation} \label{3.07}
\begin{array}{l}
\displaystyle{%
  c_V\equiv\frac{C_V}{N}=\frac{3}{2}\bigg(1-\frac{q}{2^{7/2}g}\bigg), %
}%
\end{array}
\end{equation}
\vspace{00mm}
\begin{equation} \label{3.08}
\begin{array}{l}
\displaystyle{%
  c_p\equiv\frac{C_p}{N}=\frac{5}{2}\bigg(1-\frac{3q}{2^{7/2}g}\bigg). %
}%
\end{array}
\end{equation}
The squares of the isothermal and adiabatic speeds of sound with
account of quantum corrections have the form
\begin{equation} \label{3.09}
\begin{array}{l}
\displaystyle{%
  u_T^2=\frac{T}{m}\bigg(1+\frac{q}{2^{3/2}g}\bigg), %
}%
\end{array}
\end{equation}
\vspace{00mm}
\begin{equation} \label{3.10}
\begin{array}{l}
\displaystyle{%
  u_\sigma^2=\frac{5T}{3m}\bigg(1+\frac{q}{2^{5/2}g}\bigg). %
}%
\end{array}
\end{equation}
Quantum corrections result in an increase of the speed of sound. Let
us also present the ratios of heat capacities and squares of speeds
\begin{equation} \label{3.11}
\begin{array}{l}
\displaystyle{%
  \frac{C_p}{C_V}=\frac{u_\sigma^2}{u_T^2}=\frac{5}{3}\bigg(1-\frac{q}{2^{5/2}g}\bigg), %
}%
\end{array}
\end{equation}
as well as the formula for the difference of heat capacities
\begin{equation} \label{3.12}
\begin{array}{l}
\displaystyle{%
  \frac{C_p-C_V}{N}=1-\frac{3q}{2^{5/2}g}. %
}%
\end{array}
\end{equation}
Thermodynamic coefficients (\ref{35})\,--\,(\ref{37}) with account
of the first quantum correction have the form
\begin{equation} \label{3.13}
\begin{array}{c}
\displaystyle{%
  \alpha_p=\frac{1}{T}\bigg(1-\frac{5}{2^{7/2}}\frac{q}{g}\bigg), %
}%
\end{array}
\end{equation}
\vspace{00mm}
\begin{equation} \label{3.14}
\begin{array}{c}
\displaystyle{%
  \gamma_T=\frac{1}{nT}\bigg(1-\frac{q}{2^{3/2}g}\bigg), %
}%
\end{array}
\end{equation}
\vspace{00mm}
\begin{equation} \label{3.15}
\begin{array}{c}
\displaystyle{%
  \beta_V=\frac{1}{T}\bigg(1-\frac{3}{2^{7/2}}\frac{q}{g}\bigg). %
}%
\end{array}
\end{equation}

\subsection{Fermi gas close to degeneracy}
Let us present formulas for thermodynamic functions near $T=0$ with
account of the first temperature correction. The thermodynamic
potential in the case under consideration has the form
\begin{equation} \label{3.16}
\begin{array}{l}
\displaystyle{%
  \Omega=-\frac{4g\,m^{3/2}\mu^{5/2}V}{15\sqrt{2}\,\pi^2\hbar^3}\bigg(1+\frac{5\pi^2}{8}\frac{T^2}{\mu^2}\bigg). %
}%
\end{array}
\end{equation}
The particle number density as a function of temperature and
chemical potential is given by the formula
\begin{equation} \label{3.17}
\begin{array}{l}
\displaystyle{%
  n(T,\mu)=\frac{\sqrt{2}gm^{3/2}\mu^{3/2}}{3\pi^2\hbar^3}\bigg(1+\frac{\pi^2}{8}\frac{T^2}{\mu^2}\bigg). %
}%
\end{array}
\end{equation}
From the latter formula, the chemical potential can be expressed
through the temperature and particle number density
\begin{equation} \label{3.18}
\begin{array}{l}
\displaystyle{%
  \mu=\mu_0\bigg(1-\frac{\pi^2}{12}\frac{T^2}{\mu_0^2}\bigg), %
}%
\end{array}
\end{equation}
where
\begin{equation} \label{3.19}
\begin{array}{l}
\displaystyle{%
  \mu_0\equiv\varepsilon_F=\bigg(\frac{6\pi^2}{g}\bigg)^{\!2/3}\frac{\hbar^2}{2m}\,n^{2/3} %
}%
\end{array}
\end{equation}
is the value of the chemical potential at $T=0$, coinciding with the
Fermi energy. As seen, at $T=0$ the chemical potential takes on a
finite positive value, so that at $T\rightarrow 0$ the parameter
$t=\mu/T$ increases indefinitely $t\rightarrow\infty$. The Fermi
momentum is equal to
\begin{equation} \label{3.20}
\begin{array}{l}
\displaystyle{%
  p_F\equiv\sqrt{2m\varepsilon_F}=\hbar\bigg(\frac{6\pi^2}{g}\bigg)^{\!1/3}\,n^{1/3}. %
}%
\end{array}
\end{equation}
The energy of the gas with density $n$ that is close to degeneracy
is determined by the formula
\begin{equation} \label{3.21}
\begin{array}{l}
\displaystyle{%
  E=E_0\bigg(1+\frac{5\pi^2}{12}\frac{T^2}{\mu_0^2}\bigg), %
}%
\end{array}
\end{equation}
where the energy at zero temperature
\begin{equation} \label{3.22}
\begin{array}{l}
\displaystyle{%
  E_0=\frac{3}{10}\,V\bigg(\frac{6\pi^2}{g}\bigg)^{\!2/3}\frac{\hbar^2}{m}\,n^{5/3} = \frac{3}{5}N\mu_0. %
}%
\end{array}
\end{equation}
The pressure is given by the formula
\begin{equation} \label{3.23}
\begin{array}{l}
\displaystyle{%
  p=p_0\bigg(1+\frac{5\pi^2}{12}\frac{T^2}{\mu_0^2}\bigg), %
}%
\end{array}
\end{equation}
where the pressure at zero temperature
\begin{equation} \label{3.24}
\begin{array}{l}
\displaystyle{%
  p_0=\frac{1}{5}\bigg(\frac{6\pi^2}{g}\bigg)^{\!2/3}\frac{\hbar^2}{m}\,n^{5/3} = \frac{2}{5}\,\mu_0n. %
}%
\end{array}
\end{equation}
The entropy at absolute zero turns to zero, so that the third law of
thermodynamics in the Planck formulation holds. At low temperatures
the entropy is proportional to temperature, and the next term of the
expansion is proportional to the cube of temperature
\begin{equation}
\label{3.25}
\begin{array}{l}
\displaystyle{%
  S=V\frac{m}{\hbar^2}\bigg(\frac{g\pi}{6}\bigg)^{\!2/3}n^{1/3}\bigg(T-\frac{\pi^2}{10}\frac{T^3}{\mu_0^2}\bigg)= %
  \frac{\pi^2N}{2}\bigg(\frac{T}{\mu_0}-\frac{\pi^2}{10}\frac{T^3}{\mu_0^3}\bigg). %
}%
\end{array}
\end{equation}
The heat capacities have the form
\begin{equation} \label{3.26}
\begin{array}{l}
\displaystyle{%
  \frac{C_V}{N}=\frac{\pi^2}{2}\bigg(\frac{T}{\mu_0}-\frac{3\pi^2}{10}\frac{T^3}{\mu_0^3}\bigg), %
}%
\end{array}
\end{equation}
\vspace{00mm}
\begin{equation} \label{3.27}
\begin{array}{l}
\displaystyle{%
  \frac{C_p}{N}=\frac{\pi^2}{2}\bigg(\frac{T}{\mu_0}+\frac{\pi^2}{30}\frac{T^3}{\mu_0^3}\bigg), %
}%
\end{array}
\end{equation}
and their difference is proportional to the cube of temperature
\begin{equation} \label{3.28}
\begin{array}{l}
\displaystyle{%
  \frac{C_p-C_V}{N}=\frac{\pi^4}{6}\frac{T^3}{\mu_0^3}. %
}%
\end{array}
\end{equation}
The ratio of heat capacities tends to unity as the temperature tends to zero %
\begin{equation} \label{3.29}
\begin{array}{l}
\displaystyle{%
  \frac{C_p}{C_V}=1+\frac{\pi^2}{3}\frac{T^2}{\mu_0^2}. %
}%
\end{array}
\end{equation}
The squares of the sound speeds in the low-temperature limit have the form %
\begin{equation} \label{3.30}
\begin{array}{l}
\displaystyle{%
  u_T^2=\frac{2}{3}\frac{\mu_0}{m}\bigg(1+\frac{\pi^2}{12}\frac{T^2}{\mu_0^2}\bigg), %
}%
\end{array}
\end{equation}
\vspace{00mm}
\begin{equation} \label{3.31}
\begin{array}{l}
\displaystyle{%
  u_\sigma^2=\frac{2}{3}\frac{\mu_0}{m}\bigg(1+\frac{5\pi^2}{12}\frac{T^2}{\mu_0^2}\bigg). %
}%
\end{array}
\end{equation}
Thermodynamic coefficients (\ref{35})\,--\,(\ref{37}) in the low-temperature limit are %
\begin{equation} \label{3.32}
\begin{array}{c}
\displaystyle{%
  \alpha_p=\frac{\pi^2}{2\mu_0}\bigg(\frac{T}{\mu_0}-\frac{23\pi^2}{60}\frac{T^3}{\mu_0^3}\bigg), %
}%
\end{array}
\end{equation}
\vspace{00mm}
\begin{equation} \label{3.33}
\begin{array}{c}
\displaystyle{%
  \gamma_T=\frac{3}{2}\frac{1}{n\mu_0}\bigg(1-\frac{\pi^2}{12}\frac{T^2}{\mu_0^2}\bigg), %
}%
\end{array}
\end{equation}
\vspace{00mm}
\begin{equation} \label{3.34}
\begin{array}{c}
\displaystyle{%
  \beta_V=\frac{5\pi^2}{6\mu_0}\bigg(\frac{T}{\mu_0}-\frac{43\pi^2}{60}\frac{T^3}{\mu_0^3}\bigg). %
}%
\end{array}
\end{equation}

\subsection{Fermi gas at zero chemical potential}
There occurs Bose-Einstein condensation in the ideal gas of bosons
at zero chemical potential. In the gas of fermions the chemical
potential at high temperatures is negative. As the temperature
decreases, it also turns to zero at a certain temperature. However,
this phenomenon is not associated with any phase transition. As the
temperature further decreases down to zero, the chemical potential
remains positive. Let us consider separately the state of the Fermi
gas under conditions when its chemical potential is zero. In this
case, quantum regularities remain essential. The quantumness
(\ref{14}) at $\mu=0$ and $g=2$ is not small $q_0=g\Phi_{3/2}(0)=1.53$. %
The particle number density, energy and pressure at $\mu=0$ are given by %
\begin{equation} \label{3.35}
\begin{array}{l}
\displaystyle{%
  n=\frac{g}{\Lambda^3}\,\Phi_{3/2}(0), \qquad \frac{E}{V}=\frac{3}{2}\frac{gT}{\Lambda^3}\,\Phi_{5/2}(0), \qquad %
  p=\frac{gT}{\Lambda^3}\,\Phi_{5/2}(0). %
}%
\end{array}
\end{equation}
Here are the numerical values of $\Phi_s(0)$:
$$ \Phi_{1/2}(0)=0.60490, \qquad \Phi_{3/2}(0)=0.76515, \qquad \Phi_{5/2}(0)=0.86720.   $$

Sometimes it is convenient to pass to a dimensionless notation of
thermodynamic quantities. For this purpose, let us choose an
arbitrary non-zero temperature $T_0$. The corresponding de Broglie
wavelength $\Lambda_0\equiv\Lambda(T_0)=\big(2\pi\hbar^2/mT_0\big)^{\!1/2}$. %
At this temperature the particle number density, energy and pressure are %
\begin{equation} \label{3.36}
\begin{array}{l}
\displaystyle{%
  n_0=\frac{g}{\Lambda_0^3}\,\Phi_{3/2}(0), \qquad \frac{E_0}{V}=\frac{3}{2}\frac{gT_0}{\Lambda_0^3}\,\Phi_{5/2}(0), \qquad %
  p_0=\frac{gT_0}{\Lambda_0^3}\,\Phi_{5/2}(0). %
}%
\end{array}
\end{equation}
The dimensionless particle number density, energy and pressure at
$\mu=0$ are functions of only the dimensionless temperature $\tau\equiv T/T_0$: %
\begin{equation} \label{3.37}
\begin{array}{l}
\displaystyle{%
  \tilde{n}\equiv\frac{n}{n_0}=\tau^{3/2}, \qquad \tilde{p}\equiv\frac{p}{p_0}=\tau^{5/2}, \qquad \tilde{E}\equiv\frac{E}{E_0}=\tau^{5/2}. %
}%
\end{array}
\end{equation}

\begin{figure}[b!]
\vspace{00mm} \hspace{0mm} %
\centering
\includegraphics[width = 7.02cm]{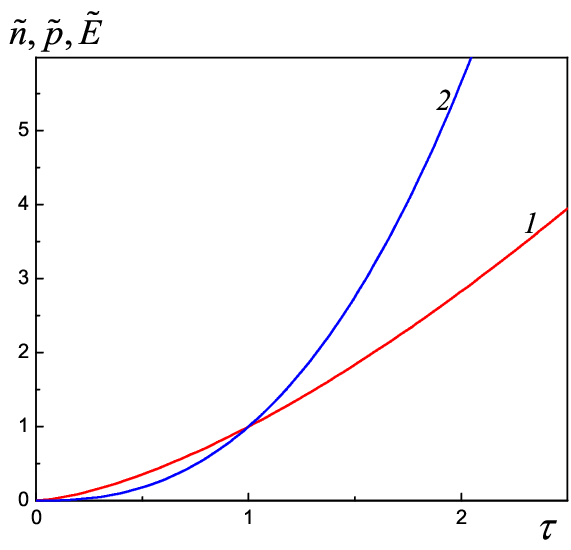} 
\vspace{-4mm} %
\caption{\label{fig07} 
Dependences of the density $\tilde{n}(\tau)$ -- {\it 1}, the
pressure and energy $\tilde{p}(\tau),\tilde{E}(\tau)$ -- {\it 2} on
the dimensionless temperature $\tau\equiv T/T_0$ at $\mu=0$.
}%
\end{figure}

These dependencies are presented in Fig.\,7. The entropy and heat
capacities per one particle are the following numbers
\begin{equation} \label{3.38}
\begin{array}{l}
\displaystyle{%
  \sigma=\frac{5}{2}\frac{\Phi_{5/2}(0)}{\Phi_{3/2}(0)}=2.833, \quad c_V=1.404, \quad c_p=2.097. %
}%
\end{array}
\end{equation}
The pressure, particle number density and temperature under
condition $\mu=0$ are linked by the relation
\begin{equation} \label{3.39}
\begin{array}{l}
\displaystyle{%
  \frac{nT}{p}=\theta=\frac{\Phi_{3/2}(0)}{\Phi_{5/2}(0)}=0.882.  %
}%
\end{array}
\end{equation}
The speeds of sound (\ref{26}),\,(\ref{27}) are proportional to temperature %
\begin{equation} \label{3.40}
\begin{array}{l}
\displaystyle{%
  u_T^2=\frac{T}{m}\frac{\Phi_{3/2}(0)}{\Phi_{1/2}(0)}, \qquad u_\sigma^2=\frac{5T}{3m}\frac{\Phi_{5/2}(0)}{\Phi_{3/2}(0)}.  %
}%
\end{array}
\end{equation}
The coefficient of volumetric expansion $\alpha_p$ (\ref{35}) and
the isochoric thermal pressure coefficient $\beta_V$ (\ref{37}) are
inversely proportional to temperature
\begin{equation} \label{3.41}
\begin{array}{l}
\displaystyle{%
  \alpha_p=\frac{5}{2T}\bigg[\frac{\Phi_{5/2}(0)\Phi_{1/2}(0)}{\Phi_{3/2}^2(0)}-\frac{3}{5}\bigg]=\frac{0.740}{T}, \quad %
  \beta_V=\frac{5}{2T}\bigg[1-\frac{3}{5}\frac{\Phi_{3/2}^2(0)}{\Phi_{5/2}(0)\Phi_{1/2}(0)}\bigg]=\frac{0.826}{T}. %
}%
\end{array}
\end{equation}
The isothermal compressibility $\gamma_T$ (\ref{36}) is given by the formula %
\begin{equation} \label{3.42}
\begin{array}{l}
\displaystyle{%
  \gamma_T=\frac{\Lambda^3}{gT}\frac{\Phi_{1/2}(0)}{\Phi_{3/2}^2(0)}= %
  \frac{1}{g}\bigg(\frac{2\pi\hbar^2}{m}\bigg)^{\!3/2}\,\frac{\Phi_{1/2}(0)}{\Phi_{3/2}^2(0)}\frac{1}{T^{5/2}}, %
}%
\end{array}
\end{equation}
so that $\gamma_T\sim T^{-5/2}$.

\section{Conclusion}\vspace{00mm} %
The paper provides a reference material containing formulas for the
basic thermodynamic quantities of the ideal Fermi gas, expressed
through the standard Fermi-Stoner functions, and considers the basic
properties of these functions. The limiting cases of high and low
temperatures and the special case of zero chemical potential are
also considered. All given relations are valid for systems of large
volume and large number of particles. Quantum distribution functions
over discrete states in systems of non-interacting fermions and
bosons in a finite volume with an arbitrary, including small, number
of particles and their properties are considered by the authors in
[13\,--\,15].

\newpage
\appendix
\section{\hspace{0mm}appendix} 

{\small\noindent The derivation of the low-temperature asymptotics
with taking into account exponential corrections.}\vspace{4mm}

Let us write the initial formula (\ref{2.1}), valid at $t>0$, in the
form $\displaystyle{\Phi_s(t)\equiv\frac{2t^s}{\Gamma(s)}\,J_s(t)}$, where %
$\displaystyle{J_s(t)\equiv\int_0^\infty\frac{x^{2s-1}dx}{e^{\,t(x^2-1)}+1}=J_s^{(1)}(t)+J_s^{(2)}(t)}$, and %
\begin{equation} \label{A01}
\begin{array}{l}
\displaystyle{%
  J_s^{(1)}(t)\equiv\int_0^1\frac{x^{2s-1}dx}{e^{\,t(x^2-1)}+1}, \qquad %
  J_s^{(2)}(t)\equiv\int_1^\infty\frac{x^{2s-1}dx}{e^{\,t(x^2-1)}+1}.  %
}%
\end{array}
\end{equation}
Making a substitution $y=1-x^2$ in the first integral and $y=x^2-1$
in the second one, we obtain
\begin{equation} \label{A02}
\begin{array}{cc}
\displaystyle{%
  J_s^{(1)}(t)=\frac{1}{2}\bigg[\frac{1}{s}+\sum_{n=1}^\infty(-1)^n\int_0^1\!dy(1-y)^{s-1}e^{-nty}\bigg], %
}\vspace{3mm}\\ %
\displaystyle{\hspace{0mm}%
  J_s^{(2)}(t)=-\frac{1}{2}\sum_{n=1}^\infty(-1)^n\int_0^\infty\!dy(1+y)^{s-1}e^{-nty}. %
}%
\end{array}
\end{equation}
Then we can write down $J_s(t)=J_s^{(3)}(t)+J_s^{(4)}(t)$, where
\begin{equation} \nonumber
\begin{array}{cc}
\displaystyle{%
  J_s^{(3)}(t)=\frac{1}{2}\bigg[\frac{1}{s}+\sum_{n=1}^\infty(-1)^nZ_{sn}(t)\bigg], \quad %
  J_s^{(4)}(t)=-\frac{1}{2}\sum_{n=1}^\infty(-1)^n{\rm B}_{sn}(t). %
}%
\end{array}
\end{equation}
Here
\begin{equation} \label{A03}
\begin{array}{l}
\displaystyle{%
  Z_{sn}(t)\equiv\int_0^1\!dy\big[(1-y)^{s-1}-(1+y)^{s-1}\big]e^{-nty},  %
}%
\end{array}
\end{equation}
\vspace{00mm}
\begin{equation} \label{A04}
\begin{array}{l}
\displaystyle{%
  {\rm B}_{sn}(t)\equiv\int_1^\infty\!dy(1+y)^{s-1}e^{-nty}.  %
}%
\end{array}
\end{equation}
To calculate the integral (\ref{A03}), one should use the expansions
$(\nu>0)$ $\big(1\pm x\big)^\nu=\sum_{r=0}^\infty\,(\pm 1)^rC_\nu^r x^r$. %
If $\nu\equiv n$ is an integer, then $C_\nu^r\equiv C_n^r=\frac{n!}{(n-r)!\,r!}$. %
If $\nu$ is a half-integer, then $C_\nu^r=\frac{\nu(\nu-1)\ldots (\nu-r+1)}{r!}$. %
For a negative exponent $\big(1\pm x\big)^{-\nu}=\sum_{r=0}^\infty\,(\pm 1)^rC_{-\nu}^r x^r$, %
where $C_{-\nu}^r\equiv(-1)^r\!\cdot\!\frac{\nu(\nu+1)\,\ldots\,(\nu+r-1)}{r!}$. Then we obtain %
\begin{equation} \label{A05}
\begin{array}{l}
\displaystyle{%
  Z_{sn}(t)=\sum_{r=1}^\infty C_{s-1}^{\,r}\big[(-1)^r-1\big]\int_0^1\!dy\,y^r e^{-nty}.  %
}%
\end{array}
\end{equation}
Here
\begin{equation} \nonumber
\begin{array}{cc}
\displaystyle{%
  \int_0^1\!dy\,y^r e^{-nty}=-e^{-nt}a_{nr}(t)+\frac{r!}{(nt)^{r+1}} %
}%
\end{array}
\end{equation}
and
\begin{equation} \label{A06}
\begin{array}{l}
\displaystyle{%
  a_{nr}(t)\equiv\sum_{l=1}^{r+1}\frac{r!}{(nt)^l(r-l+1)!}=  %
  \frac{1}{nt}+\frac{r}{(nt)^2}+\frac{r(r-1)}{(nt)^3}+\ldots +\frac{r!}{(nt)^r}+\frac{r!}{(nt)^{r+1}}. %
}%
\end{array}
\end{equation}
Therefore
\begin{equation}\nonumber
\begin{array}{l}
\displaystyle{%
  Z_{sn}(t)=\sum_{r=1}^\infty C_{s-1}^{\,r}\big[(-1)^r-1\big]\! %
  \bigg[-e^{-nt}a_{nr}(t)+\frac{r!}{(nt)^{r+1}}\bigg].
}%
\end{array}
\end{equation}
As a result, taking into account that terms only with odd $r=2k+1$
remain in the sum, we have
\begin{equation} \label{A07}
\begin{array}{l}
\displaystyle{%
  J_s^{(3)}(t)=\frac{1}{2}\left\{\frac{1}{s}+2\sum_{k=0}^\infty C_{s-1}^{2k+1}\! %
  \left[\sum_{n=1}^\infty(-1)^ne^{-nt}a_{n,2k+1}(t)-\frac{(2k+1)!}{t^{2k+2}}\sum_{n=1}^\infty\frac{(-1)^n}{n^{2k+2}}\right]\!\right\}.  %
}%
\end{array}
\end{equation}
For the zeta function $\displaystyle{\zeta(z)=\sum_{k=1}^\infty\frac{1}{k^z} }$\, $({\rm Re}\,z > 1)$, %
there holds the relation $\displaystyle{\sum_{k=1}^\infty\frac{(-1)^k}{k^z}=\big(2^{1-z}-1\big)\zeta(z) }$, so that %
\begin{equation} \label{A08}
\begin{array}{l}
\displaystyle{%
  J_s^{(3)}(t)=\frac{1}{2}\left\{\frac{1}{s}+2\sum_{k=0}^\infty C_{s-1}^{2k+1} %
  \bigg[A_k(t)+\frac{(2k+1)!}{t^{2k+2}}\big(1-2^{-(2k+1)}\big)\zeta(2k+2)\bigg]\right\},  %
}%
\end{array}
\end{equation}
where the notation is used
\begin{equation} \nonumber
\begin{array}{cc}
\displaystyle{%
  A_k(t)\equiv\sum_{n=1}^\infty(-1)^ne^{-nt}a_{n,2k+1}(t)= %
  (2k+1)!\sum_{l=1}^{2k+2}\frac{1}{(2k+2-l)!\,t^l}\sum_{n=1}^\infty\frac{(-1)^ne^{-nt}}{n^l}= %
}\vspace{3mm}\\ %
\displaystyle{\hspace{0mm}%
  =-(2k+1)!\sum_{l=1}^{2k+2}\frac{1}{(2k+2-l)!}\frac{\Phi_l(-t)}{t^l}. %
}%
\end{array}
\end{equation}

Calculation of the quantity (\ref{A04}) gives $\displaystyle{{\rm B}_{sn}(t)=\frac{e^{nt}}{(nt)^s}\,I_s(2nt) }$, where %
\begin{equation} \label{A09}
\begin{array}{l}
\displaystyle{%
  I_s(p)=\int_p^\infty\!du\,u^{s-1}e^{-u}\equiv \Gamma(s,p), %
}%
\end{array}
\end{equation}
and $\Gamma(s,p)$ -- the incomplete gamma function \cite{AS}. \newline %
Two cases should be considered separately: 1) $s$ -- an integer, 2) $s$ -- a half-integer. %
\newline\noindent %
1) $s$ -- an integer:
\begin{equation} \label{A10}
\begin{array}{l}
\displaystyle{\hspace{-3mm}%
  I_s(p)=e^{-p}\big[p^{s-1}+(s-1)p^{s-2}+\ldots+ (s-1)!\,p +(s-1)!\big]= %
  e^{-p}\sum_{r=0}^{s-1}\frac{(s-1)!}{(s-r-1)!}\,p^{s-1-r}. %
}%
\end{array}
\end{equation}
\vspace{-2mm}\newline\noindent %
2) $s$ -- a half-integer. \newline\noindent %
If $s=1/2$, then
\begin{equation} \label{A11}
\begin{array}{l}
\displaystyle{%
  I_{1/2}(p)\equiv\int_p^\infty\!du\frac{e^{-u}}{\sqrt{u}}=\sqrt{\pi}\,{\rm erfc}\big(\sqrt{p}\big), %
}%
\end{array}
\end{equation}
where the complementary probability integral:
\begin{equation} \nonumber
\begin{array}{l}
\displaystyle{%
  {\rm erfc}(z)\equiv\frac{2}{\sqrt{\pi}}\int_z^\infty e^{-t^2}dt. %
}%
\end{array}
\end{equation}
In particular, for $x\gg 1$ the asymptotics holds\, $\displaystyle{{\rm erfc}(z)\approx\frac{e^{-z^2}}{\sqrt{\pi}\,z} }$. %
\newline\noindent %
For $s>1/2$, %
\begin{equation} \label{A12}
\begin{array}{l}
\displaystyle{%
  I_{s}(p)=e^{-p}\sum_{r=0}^{s-3/2}\frac{(s-1)(s-2)\ldots 1/2}{(s-r-1)(s-r-2)\ldots 1/2}\, %
  p^{s-1-r} + (s-1)(s-2)\ldots\frac{3}{2}\!\cdot\!\frac{1}{2}\!\cdot\!I_{1/2}(p).  %
}%
\end{array}
\end{equation}
In particular,
\begin{equation} \label{A13}
\begin{array}{l}
\displaystyle{%
  I_{3/2}(p)=e^{-p}\!\cdot\!p^{1/2} + \frac{1}{2}\,I_{1/2}(p), \quad %
  I_{5/2}(p)=e^{-p}\bigg(p^{3/2} + \frac{3}{2}\,p^{1/2}\bigg)+\frac{3}{2}\!\cdot\!\frac{1}{2}\!\cdot\!I_{1/2}(p). %
}%
\end{array}
\end{equation}

Thus, the integral $J_s(t)$ is decomposed into the sum of the main
$J_s^{(m)}(t)$ and exponential $J_s^{({\rm exp})}(t)$ parts
\begin{equation} \label{A14}
\begin{array}{l}
\displaystyle{%
  J_s(t)=J_s^{(m)}(t)+J_s^{({\rm exp})}(t). %
}%
\end{array}
\end{equation}
The main part of the decomposition:
\begin{equation} \label{A15}
\begin{array}{l}
\displaystyle{%
  J_s^{(m)}(t)=\frac{1}{2s}+\sum_{k=0}^\infty C_{s-1}^{2k+1}\frac{(2k+1)!}{t^{2k+2}}\big(1-2^{-(2k+1)}\big)\zeta(2k+2). %
}%
\end{array}
\end{equation}
The exponential part of the decomposition:
\begin{equation} \label{A16}
\begin{array}{l}
\displaystyle{%
  J_s^{({\rm exp})}(t)\equiv J_s^{({\rm exp1})}(t)+J_s^{({\rm exp2})}(t), %
}%
\end{array}
\end{equation}
where
\begin{equation} \label{A17}
\begin{array}{l}
\displaystyle{%
  J_s^{({\rm exp1})}(t)=\sum_{k=0}^\infty C_{s-1}^{2k+1}A_k(t)= %
  -\sum_{k=0}^\infty C_{s-1}^{2k+1}(2k+1)!\sum_{l=1}^{2k+2}\frac{1}{(2k+2-l)!}\frac{\Phi_l(-t)}{t^l}, %
}%
\end{array}
\end{equation}
\vspace{00mm}
\begin{equation} \label{A18}
\begin{array}{l}
\displaystyle{%
  J_s^{({\rm exp2})}(t)=\frac{1}{2}\sum_{n=1}^\infty (-1)^{n+1}{\rm B}_{sn}(t). %
}%
\end{array}
\end{equation}
At an integer $s$:
\begin{equation} \label{A19}
\begin{array}{cc}
\displaystyle{%
  J_s^{({\rm exp2})}(t)=-\frac{1}{2}(s-1)!\sum_{r=0}^{s-1}\frac{1}{2^{1-s+r}(s-1-r)!\,t^{r+1}} %
  \sum_{n=1}^\infty (-1)^n\frac{e^{-nt}}{n^{r+1}}= %
}\vspace{3mm}\\ %
\displaystyle{\hspace{0mm}%
  =\frac{1}{2}(s-1)!\sum_{r=0}^{s-1}\frac{1}{2^{1-s+r}(s-1-r)!}\frac{\Phi_{r+1}(-t)}{t^{r+1}}. %
}%
\end{array}
\end{equation}
At a half-integer $\displaystyle{s=\frac{3}{2},\,\frac{5}{2},\,\ldots\,}$: %
\begin{equation} \label{A20}
\begin{array}{cc}
\displaystyle{%
  J_s^{({\rm exp2})}(t)=\frac{1}{2}\sum_{r=0}^{s-3/2}\frac{(s-1)(s-2)\ldots 1/2}{(s-r-1)(s-r-2)\ldots 1/2}%
  \!\cdot\!2^{s-1-r}\!\cdot\!\frac{\Phi_{r+1}(-t)}{t^{r+1}}+ %
}\vspace{3mm}\\ %
\displaystyle{\hspace{09mm}%
  +\frac{1}{2}(s-1)(s-2)\ldots\frac{3}{2}\!\cdot\!\frac{1}{2}\,\frac{1}{t^s} %
  \sum_{n=1}^\infty(-1)^{n+1}\frac{e^{nt}}{n^s}\,I_{1/2}(2nt). %
}%
\end{array}
\end{equation}
For $\displaystyle{s=\frac{1}{2}}$, %
\begin{equation} \label{A21}
\begin{array}{l}
\displaystyle{%
  J_{1/2}^{({\rm exp2})}(t)\equiv -\frac{1}{2}\,\frac{1}{t^{1/2}}\sum_{n=1}^\infty (-1)^n\frac{e^{nt}}{n^{1/2}}\,%
  \int_{2nt}^\infty du\frac{e^{-u}}{\sqrt{u}}\,. %
}%
\end{array}
\end{equation}

\newpage\noindent\vspace{02mm} %
The special cases for $\displaystyle{\Phi_s(t)=\frac{2t^s}{\Gamma(s)}\,J_s(t)}$: %

\noindent\vspace{00mm} %
$\displaystyle{1) \,s=\frac{1}{2}; \,\Phi_{1\!/2}(t)=\frac{2t^{1\!/2}}{\Gamma(1/2)}\,J_{1\!/2}(t). }$ %
\begin{equation} \nonumber
\begin{array}{c}
\displaystyle{%
  J_{1/2}^{(m)}=1-\frac{\pi^2}{24}\!\cdot\!\frac{1}{t^2}-\frac{7\pi^4}{384}\!\cdot\!\frac{1}{t^4}-\frac{31\pi^6}{1024}\!\cdot\!\frac{1}{t^6}, %
}%
\end{array}
\end{equation}
\vspace{00mm}
\begin{equation} \nonumber
\begin{array}{c}
\displaystyle{%
  J_{1/2}^{({\rm exp1})}(t)=\sum_{k=0}^\infty\frac{1\cdot 3\cdot\ldots\cdot (4k+1)}{2^{2k+1}} %
  \sum_{l=1}^{2k+2}\frac{1}{(2k+2-l)!}\frac{\Phi_l(-t)}{t^l},
}%
\end{array}
\end{equation}
\vspace{00mm}
\begin{equation} \nonumber
\begin{array}{l}
\displaystyle{%
  J_{1/2}^{({\rm exp2})}(t)\equiv -\frac{1}{2}\,\frac{1}{t^{1/2}}\sum_{n=1}^\infty (-1)^n\frac{e^{nt}}{n^{1/2}}\,%
  \int_{2nt}^\infty du\frac{e^{-u}}{\sqrt{u}}\,. %
}%
\end{array}
\end{equation}

\vspace{02mm}\noindent %
$\displaystyle{ 2) \,s=\frac{3}{2}; \,\Phi_{3/2}(t)=\frac{2t^{3/2}}{\Gamma(3/2)}\,J_{3/2}(t). }$ %
\begin{equation} \nonumber
\begin{array}{c}
\displaystyle{%
  J_{3/2}^{(m)}=\frac{1}{3}\left(1+\frac{\pi^2}{8}\!\cdot\!\frac{1}{t^2}+\frac{7\pi^4}{640}\!\cdot\!\frac{1}{t^4}+\frac{31\pi^6}{3072}\!\cdot\!\frac{1}{t^6}\right), %
}%
\end{array}
\end{equation}
\vspace{00mm}
\begin{equation} \nonumber
\begin{array}{c}
\displaystyle{%
  J_{3/2}^{({\rm exp1})}(t)=-\sum_{k=0}^\infty\frac{1\cdot(-1)\cdot(-3)\cdot\ldots\cdot(1-4k)}{2^{2k+1}} %
  \sum_{l=1}^{2k+2}\frac{1}{(2k+2-l)!}\frac{\Phi_l(-t)}{t^l},
}%
\end{array}
\end{equation}
\vspace{00mm}
\begin{equation} \nonumber
\begin{array}{l}
\displaystyle{%
  J_{3/2}^{({\rm exp2})}(t)=\frac{1}{2^{1/2}}\frac{\Phi_1(-t)}{t} +%
  \frac{1}{2^2}\frac{1}{t^{3/2}}\sum_{n=1}^\infty (-1)^{n+1}\frac{e^{nt}}{n^{3/2}}\,I_{1/2}(2nt). %
}%
\end{array}
\end{equation}

\vspace{02mm}\noindent %
$\displaystyle{ 3) \,s=\frac{5}{2}; \,\Phi_{5/2}(t)=\frac{2t^{5/2}}{\Gamma(5/2)}\,J_{5/2}(t). }$ %
\begin{equation} \nonumber
\begin{array}{c}
\displaystyle{%
  J_{5/2}^{(m)}=\frac{1}{5}\left(1+\frac{5\pi^2}{8}\!\cdot\!\frac{1}{t^2}-\frac{7\pi^4}{384}\!\cdot\!\frac{1}{t^4}-\frac{155\pi^6}{21504}\!\cdot\!\frac{1}{t^6}\right), %
}%
\end{array}
\end{equation}
\vspace{00mm}
\begin{equation} \nonumber
\begin{array}{c}
\displaystyle{%
  J_{5/2}^{({\rm exp1})}(t)=-\sum_{k=0}^\infty\frac{3\cdot1\cdot(-1)\cdot\ldots\cdot(3-4k)}{2^{2k+1}} %
  \sum_{l=1}^{2k+2}\frac{1}{(2k+2-l)!}\frac{\Phi_l(-t)}{t^l},
}%
\end{array}
\end{equation}
\vspace{00mm}
\begin{equation} \nonumber
\begin{array}{l}
\displaystyle{%
  J_{5/2}^{({\rm exp2})}(t)=\frac{1}{2}\bigg\{2^{3/2}\frac{\Phi_1(-t)}{t}+\frac{3}{2^{1/2}}\frac{\Phi_2(-t)}{t^2}\bigg\} + %
  \frac{3}{2^3}\frac{1}{t^{5/2}}\sum_{n=1}^\infty (-1)^{n+1}\frac{e^{nt}}{n^{5/2}}\,I_{1/2}(2nt). %
}%
\end{array}
\end{equation}

\newpage
\section{\hspace{0mm}appendix} 

{\small\noindent Some mathematical relations with the Riemann zeta
function and the gamma function. }\vspace{4mm}

{\it The Riemann zeta function} at ${\rm Re}\,z > 1$ is defined by the series %
\begin{equation} \label{B01}
\begin{array}{l}
\displaystyle{%
  \zeta(z)=\sum_{n=1}^\infty \frac{1}{n^z}\,. %
}%
\end{array}
\end{equation}
At ${\rm Re}\,z \le 1$, $z\neq 1$, $\zeta(z)$ is defined as the
analytic continuation of this series. At the point $z=1$ it has a pole %
$\displaystyle{ \lim_{z\rightarrow 1}\!\left[\zeta(z)-\frac{1}{z-1}\right]={\rm C}=0.577215 }$ %
-- the Euler's constant. %
\newline\noindent %
At ${\rm Re}\,z > 0$ there holds the representation
\begin{equation} \label{B02}
\begin{array}{l}
\displaystyle{%
  \big(1-2^{1-z}\big)\zeta(z)=1-\frac{1}{2^z}+\frac{1}{3^z}-\frac{1}{4^z}+\ldots =\frac{1}{\Gamma(z)}\int_0^\infty\frac{t^{z-1}}{e^t+1}\,dt\,. %
}%
\end{array}
\end{equation}
Special values: \vspace{-1mm}%
\begin{equation} \nonumber
\begin{array}{cccc}
\displaystyle{%
  \zeta(2)=\frac{\pi^2}{6}\approx 1.6449, \quad \zeta(4)=\frac{\pi^4}{90}\approx 1.0823,%
}\vspace{3mm}\\ %
\displaystyle{\hspace{00mm}%
  \zeta(6)=\frac{\pi^6}{945}\approx 1.0173, \quad \zeta(8)=\frac{\pi^8}{9450}\approx 1.0041;%
}\vspace{3mm}\\ %
\displaystyle{\hspace{00mm}%
  \zeta(3)\approx 1.202, \quad \zeta(5)\approx 1.0369, \quad \zeta(7)\approx 1.00835; %
}\vspace{3mm}\\ %
\displaystyle{\hspace{00mm}%
  \zeta(0)=-\frac{1}{2}, \quad \zeta'(0)=-\frac{1}{2}\ln(2\pi), \quad \zeta(-2n)=0; %
}\vspace{3mm}\\ %
\displaystyle{\hspace{00mm}%
  \zeta(-1)=-\frac{1}{12}\approx -0.08333, \quad \zeta(-3)=\frac{1}{120}\approx 0.008333, \quad \zeta(-5)=-\frac{1}{252}\approx -0.003968. %
}%
\end{array}
\end{equation}

{\it The gamma function} is defined as the integral
\begin{equation} \label{B03}
\begin{array}{l}
\displaystyle{%
  \Gamma(n)=\int_0^\infty x^{n-1}e^{-x}dx =\int_0^1\!\left(\ln\frac{1}{x}\right)^{\!n-1}dx, %
}%
\end{array}
\end{equation}
which has a finite value for real $n>0$.

\vspace{1mm}\noindent %
Some relations:
\begin{equation} \nonumber
\begin{array}{cc}
\displaystyle{%
  \Gamma(n+1)=n\Gamma(n), %
}\vspace{3mm}\\ %
\displaystyle{\hspace{00mm}%
  \Gamma(n)=(n-1)! \,\,\,\, (n\,-\,a\hspace{1.5mm}positive\hspace{1.5mm}integer). %
}%
\end{array}
\end{equation}

\noindent %
Special values:
\begin{equation} \nonumber
\begin{array}{cc}
\displaystyle{%
  \Gamma(1)=\Gamma(2)=1, %
}\vspace{3mm}\\ %
\displaystyle{\hspace{00mm}%
  \Gamma\bigg(\frac{1}{2}\bigg)=\sqrt{\pi}, \quad \Gamma\bigg(\frac{3}{2}\bigg)=\frac{\sqrt{\pi}}{2}, \quad \Gamma\bigg(\frac{5}{2}\bigg)=\frac{3\sqrt{\pi}}{4}, %
}\vspace{3mm}\\ %
\displaystyle{\hspace{00mm}%
  \Gamma\bigg(n+\frac{1}{2}\bigg)=1\!\cdot\!3\!\cdot\!5\!\cdot\ldots (2n-3)(2n-1)\frac{\sqrt{\pi}}{2^n}\,. %
}%
\end{array}
\end{equation}


\addcontentsline{toc}{section}{References}\include{bib} %

\newpage
\section*{List of designations}
\addcontentsline{toc}{section}{List of designations} %
\vspace{-1mm} %
\noindent %
$m$ -- the mass of a gas particle, \vspace{2mm} %

\noindent %
$k$ -- the wave number of a particle, $\hbar k$ -- its momentum, \vspace{2mm} %

\noindent %
$T$ -- the temperature in energy units, \vspace{2mm} %

\noindent %
$\beta=1/T$ -- the inverse temperature, \vspace{2mm} %

\noindent %
$g=2s+1$, $s$ -- the spin of a gas particle, \vspace{2mm} %

\noindent %
$\Lambda=\big(2\pi\hbar^2/mT\big)^{\!1/2}$ -- the thermal wavelength of a particle, \vspace{2mm} %

\noindent %
$\Lambda_0=\big(2\pi\hbar^2/mT_0\big)^{\!1/2}$ -- the thermal wavelength at some characteristic temperature $T_0$, \vspace{2mm} %

\noindent %
$\mu$ -- the chemical potential, \vspace{2mm} %

\noindent %
$f_k=\big(\exp\beta\xi_k + 1\big)^{-1}$ -- the distribution function of Fermi particles in an ideal gas, \vspace{2mm} %

\noindent %
$\xi_k=\hbar^2k^2/2m - \mu$, \vspace{2mm} %

\noindent %
$\Phi_s(t)$ -- the special functions of order $s$ (an integer or
half\,-\,integer positive number), \newline through which
thermodynamic quantities of the ideal Fermi gas are expressed, %
\vspace{2mm} %

\noindent %
$t=\mu/T$ -- the variable on which the functions $\Phi_s(t)$ depend -- the adiabaticity parameter, \vspace{2mm} %

\noindent %
$\Omega$ -- the thermodynamic potential, \vspace{2mm} %

\noindent %
$N$ -- the number of particles, \vspace{2mm} %

\noindent %
$V$ -- the volume, \vspace{2mm} %

\noindent %
$n=N/V$ -- the particle number density, \vspace{2mm} %

\noindent %
$p$ -- the pressure, \vspace{2mm} %

\noindent %
$E$ -- the energy, \vspace{2mm} %

\noindent %
$S$ -- the entropy, \vspace{2mm} %

\noindent %
$\sigma=S/N$ -- the entropy per one particle, \vspace{2mm} %

\noindent %
$C_V,\,C_p$ -- the heat capacities at constant volume and pressure, \vspace{2mm} %

\noindent %
$c_V=C_V/N,\,c_p=C_p/N$ -- the heat capacities per one particle at constant volume and pressure, \vspace{2mm} %

\noindent %
$u_T,\,u_\sigma$ -- the isothermal and adiabatic speeds of sound, \vspace{2mm} %

\noindent %
$\displaystyle{T_0=\mu_0=\varepsilon_F=\bigg(\frac{6\pi^2}{g}\bigg)^{\!\!2/3}}\frac{\hbar^2n^{2/3}}{2m}$, %
$T_0$ -- the degeneracy temperature, \newline $\mu_0$ -- the chemical potential at zero temperature, %
$\varepsilon_F$ -- the Fermi energy, \vspace{2mm} %

\noindent %
$l=n^{-1/3}$ -- the average distance between gas particles, \vspace{2mm} %

\noindent %
$q=(\Lambda/l)^3=n\Lambda^3$ -- the parameter, characterizing the
degree of ``quantumness'' \newline of the gas (quantumness),

\vspace{1.5mm}
\noindent  %
$\theta\equiv nT/p$. %

\newpage
\section*{Some fundamental constants}
\addcontentsline{toc}{section}{Some fundamental constants} %
\vspace{-1mm} %
\noindent %
The Planck constant $(\,\hbar\,)$ -- $1.055\cdot 10^{-27}$\,\,$\rm{erg\cdot s}$ \vspace{2mm} %

\noindent %
The Boltzmann constant $(\,k_B\,)$ -- $1.381\cdot 10^{-16}$\,\,$\rm{erg\cdot K^{-1}}$ \vspace{2mm} %


\noindent %
The electron mass  $(\,m_e\,)$ -- $9.110\cdot 10^{-28}$\,\,g \vspace{2mm} %

\noindent %
The proton mass  $(\,m_p\,)$ -- $1.673\cdot 10^{-24}$\,\,g \vspace{2mm} %

\noindent %
The mass of $^3$He atom $(\,m_{\,^3{\rm He}}\,)$ -- $5.01\cdot 10^{-24}$\,\,g \vspace{2mm} %

\noindent %
The electron charge $(\,e\,)$ -- $4.803\cdot 10^{-10}$\,\,cm$^{3\!/2}$g$^{1\!/2}$s$^{-1}$  

\end{document}